\documentstyle[11pt]{article}

\newcommand{\vs}[1]{\rule[- #1 mm]{0mm}{#1 mm}}
\newcommand{\beq}{\begin{equation}}
\newcommand{\eeq}{\end{equation}}
\newcommand{\beql}{\begin{eqnarray}}
\newcommand{\eeql}{\end{eqnarray}}

\newcommand{\lp}{\left(}
\newcommand{\rp}{\right)}
\newcommand{\scr}{\scriptstyle}

\newcommand{\om}{\omega}
\newcommand{\al}{\alpha}
\newcommand{\lam}{\lambda}
\newcommand{\Lam}{\Lambda}
\newcommand{\eps}{\epsilon}

\newcommand{\dV}{\frac{d}{dV}}
\newcommand{\dxi}{\frac{dx_i}{dV}(p)}
\newcommand{\parV}{\frac{\partial}{\partial V}}
\newcommand{\cI}{\oint_{\cal C}\frac{d\om}{2\pi i}}
\newcommand{\bI}{\int\limits_{x_3}^{x_2}}
\newcommand{\Vp}{V^{\prime}}

\newcommand{\pho}{\phi^{(0)}}

\newcommand{\sect}[1]{\setcounter{equation}{0}\section{#1}}

\begin{document}

\begin{titlepage}

\rightline{\Large {NBI-HE\ -96-30}}
\rightline{\Large {ITP-UH-08/96}}
\rightline{\Large {hep-th/9606004}}

\vs{10}

\begin{center}

{\LARGE {\bf Higher genus correlators for the hermitian\\[.5cm]
             matrix model with multiple cuts}}\\[2cm]

{\Large G. Akemann}\\[.5cm]
{\em Niels Bohr Institute\\
Blegdamsvej 17, DK-2100 Copenhagen {\O}, Denmark}\\
{akemann@nbi.dk}\\
{\em and\\
Institut f\"ur Theoretische Physik, Universit\"at Hannover\\
Appelstra{\ss}e 2, 30167 Hannover, Germany}\\
{akemann@itp.uni-hannover.de}\\[.5cm]

\end{center} 

\vs{10}

\centerline{ {\bf Abstract}}

An iterative scheme is set up for solving the loop equation 
of the hermitian one-matrix model with a multi-cut structure.
Explicit results are presented for genus one for an arbitrary
but finite number of cuts. Due to the complicated form of the 
boundary conditions, the loop correlators now contain elliptic 
integrals. This demonstrates the existence of new universality 
classes for the hermitian matrix model. The two-cut solution 
is investigated in more detail, including the double-scaling 
limit. It is shown, that in special cases it differs from the 
known continuum solution with one cut.

\end{titlepage}

\renewcommand{\thefootnote}{\arabic{footnote}}
\setcounter{footnote}{0}

\sect{Introduction}\label{intro}

\indent

The study of multi-cut solutions of matrix models is interesting for several 
reasons. First of all it completes the perturbative analysis around $N=\infty$,
exhausting the full parameter space of the coupling constants for all possible
kinds of solutions. This is important in order to tell, whether there exist 
different types of phase transitions, universality classes or continuum limits
than the ones, which are already known for the one-cut solution of the model.

Secondly there may be applications apart from quantum gravity to two 
dimensional QCD or statistical physics. Results of random matrix theory
are used for example in the study of the spectrum of the QCD inspired
discretised Dirac operator. The merging of several cuts
into a one-cut solution of a complex matrix model with ``temperature''
would be of interest there in various limits (\cite{JNZ96} and references
therein).

In this paper the approach of iteratively solving the loop equation for the
hermitian one-matrix model \cite{AMB93} is chosen and generalised to an
arbitrary but finite number of cuts. The success of this method is based on
a nonperturbative treatment in the coupling constants, which is 
valid for an arbitrary polynomial potential. Even away from the double
scaling limit higher orders in the genus expansion
can be obtained explicitly. The validity
for an arbitrary potential makes it possible
to read off immediately, whether the results are universal or not. Here this
will lead to a whole set of new universality classes for the planar
two-point function or two-loop correlator. 

Another reason for choosing the loop equation techniques is that it seems,
that the method of orthogonal polynomials is not applicable to multi-cut 
solutions in general. Solving the string equation for the norms of the 
polynomials
numerically, instabilities or a so-called chaotic behaviour has been found
by a variety of authors \cite{LE92,JU91,SASU91,SENE92,BDJT93}. The region,
where these instabilities occur, seems to match precisely with the part of the
phase space, where the multi-cut solutions dominate. This has been shown
for the hermitian model with an even sextic potential as an example
\cite{JU91,GAPHD}. So apart from the symmetric two-cut solution and some 
cases of degenerate minima of the potential, where smooth numerical solutions
have been found \cite{LE92,SENE92,GAPHD}, an analytic expression for the
orthogonal polynomials even for the planar limit is still missing in the case
of a generic multi-cut situation\footnote{The two-component ansatz for the 
recursion coefficients made in \cite{DE90}, \cite{CM91} and \cite{CDM92} 
is restricted to the purely symmetric case of two cuts.
The failure of a more general ansatz for more cuts is studied in \cite{LE91}.}.
As an example the nonsymmetric two-cut solution, which is presented here most 
explicitly, explores such a region of chaotic numerical solutions 
\cite{BDJT93}.

The paper is divided up into two parts. In the first part the iterative 
scheme for solving the model with $s$ cuts is presented, including 
results for genus one. The second part is then devoted to a more 
detailed study of the two-cut solution, which also contains an analysis of
the double-scaling limit.

In section 2 and 3 the main definitions and the loop equations for the 
hermitian model with multiple cuts are displayed, following closely the 
notation of \cite{AMB93}. The planar solution is derived in section 4, 
including the required boundary conditions. Then, in section 5 the iterative
scheme for higher genera is presented. Special care has to be taken to the
inversion of the loop equation due to the zero modes, which occur for
more than one cut. The one-loop correlator of genus one is given for an
arbitrary number of cuts.

Turning to the two-cut solution, section 6 contains the full two-point
function as well as the free energy of genus one. After
a short discussion of general properties in the symmetric limit, 
section 7 deals with the different
possibilities of taking a continuum limit, displaying some new results.
Section 8 closes with a final discussion and future prospects.

\sect{Basic definitions} \label{basic}

\indent

Throughout the paper the same notation as in \cite{AMB93} is used,
which is redisplayed here for completeness.
The partition function of the hermitian one-matrix model is 
defined by
\beq
Z \ [N,\{g_i\}] \ \equiv \ e^{N^2 F[N,\{g_i\}]} \ \equiv \
\int d\phi \ \mbox{exp}(-N \ \mbox{Tr} V(\phi)) \ , \label{Z}
\eeq
where the integration is over hermitian $N\times N$ matrices $\phi$.
The matrix potential is given by the following power series,
\beq
V(\phi) \ \equiv \ \sum_{j=1}^\infty \frac{g_j}{j}\phi^j \ .
\eeq
A specific potential of finite order can be inspected by setting 
the extra couplings to zero in the final result.
In this way of keeping all the coupling constants $g_i$ 
throughout the calculation, they can be used as sources
for expectation values\footnote{Averages are defined as usual by
$\langle Q(\phi) \rangle \ = \ \frac{1}{Z} \int d\phi \ Q(\phi)\ 
                      \mbox{exp}(-N \ \mbox{Tr} V(\phi)) \ .$}
of the following type,
\beq
-m\frac{d}{dg_m}F \ [N,\{g_i\}] \ = \ \frac{1}{N} 
              \langle \mbox{Tr} \phi^m \rangle\ , \ \ m \in \mbox{N}_+ \ .
\eeq
Introducing the loop insertion operator
\beq
\dV(p) \ \equiv \ - \sum_{j=1}^\infty \frac{j}{p^{j+1}} \frac{d}{dg_j} \ ,
\label{dV}
\eeq
the generating functional or one-loop average $W(p)$ can thus
be obtained from the free energy $F$ 
\beql
W(p) \ &\equiv& \ \frac{1}{N} \sum_{k=0}^\infty 
             \frac{\langle \mbox{Tr}\phi^k \rangle}{p^{k+1}} 
  \ = \ \frac{1}{N} \left\langle \mbox{Tr} \frac{1}{p-\phi} \right\rangle 
  \nonumber \\ 
    \ &=& \ \dV (p) F + \frac{1}{p} \ . \label{W}
\eeql
In the same way all the multi-loop correlators can be derived by 
applying $\dV \scr (p)$ to $F$ (or to $W(p)$)
\beql
W(p_1,\ldots,p_n) \ &\equiv& \ N^{n-2} \left\langle \mbox{Tr} 
       \frac{1}{p_1-\phi}
    \cdots \mbox{Tr} \frac{1}{p_n-\phi}\right\rangle_{conn}  \nonumber \\
&=&\ \dV(p_n)\dV(p_{n-1})\cdots\dV(p_1) F \ ,
       \  \ n \ge 2 . \label{Wdef}
\eeql
Here $conn$ refers to the connected part. 
As the loop correlators and the free energy have the same genus expansion,
\beql
F &=&  \sum_{g=0}^\infty \frac{1}{N^{2g}} F_g \ ,\\
W(p_1,\ldots,p_n) &=& \sum_{g=0}^\infty \frac{1}{N^{2g}} W_g(p_1,\ldots,p_n)
\ , \label{Wg}
\eeql
eq. (\ref{Wdef}) is valid for each genus $g\geq 0$ separately.
Finally the asymptotic behaviour of $W(p)$ for large $p$ can be read off
from the definition (\ref{W})
\beq
\lim_{p\to\pm\infty}W(p) \ \sim \ \frac{1}{p} \ . \label{Wass}
\eeq
The r.h.s. does not depend on $N$, so the leading contribution clearly comes 
from the planar part
\beql
\lim_{p\to\pm\infty}W_0(p) \ &\sim& \ \frac{1}{p} \ , \nonumber\\
\lim_{p\to\pm\infty}W_g(p) \ &\sim& \ {\cal O}(\frac{1}{p^2})\ ,\ g\geq 1 \ . 
\label{Winf}
\eeql
Using this fact the genus expansion of eq. (\ref{W}) reads
\beq
W_g(p) \ =\ \dV(p)F_g \ ,\ \ g\ge1 \ ,\label{dWg}
\eeq
so the $p$-dependence of $W_g(p)$ is completely absorbed in the total
derivative. The last two equations will become important, when the genus
expanded loop equation is inverted, which determines $W(p)$ iteratively
in genus.

\sect{The loop equation}\label{theloop}

\indent

\begin{figure}[h]
\unitlength1cm
\begin{picture}(12.2,5.5)
\linethickness{1.0mm}
\multiput(2.1,2.3)(6,0){2}{\line(1,0){2.4}}
\thinlines
\put(0,2.3){\line(1,0){1.9}}
\put(4.7,2.3){\line(1,0){1}}
\multiput(6,2.3)(0.3,0){3}{\circle*{0.05}}
\put(6.9,2.3){\line(1,0){1}}
\put(10.7,2.3){\vector(1,0){2}}
\multiput(2,2.3)(2.6,0){2}{\circle*{0.2}}
\multiput(8,2.3)(2.6,0){2}{\circle*{0.2}}
\put(5.8,3.5){\circle*{0.2}}
\put(6.1,3.7){$p$}
\put(1.9,2){$x_{2s}$}
\put(4.5,2){$x_{2s-1}$}
\put(8,2){$x_2$}
\put(10.6,2){$x_1$}
\multiput(2.,2.3)(6,0){2}{\oval(1.0,2.3)[l]}
\multiput(4.6,2.3)(6,0){2}{\oval(1.0,2.3)[r]}
\multiput(2.,3.45)(6,0){2}{\line(1,0){1.2}}
\multiput(3.4,1.15)(6,0){2}{\line(1,0){1.2}}
\multiput(4.6,3.45)(6,0){2}{\vector(-1,0){1.4}}
\multiput(2.,1.15)(6,0){2}{\vector(1,0){1.4}}
\put(1.2,3.3){${\cal C}_{s}$} 
\put(7.2,3.3){${\cal C}_{1}$} 
\end{picture} 
\caption{The contour of integration 
         $ {\cal C} =  \cup_{i=1}^s {\cal C}_{i} $ }
                     \label{fig1}
\end{figure}
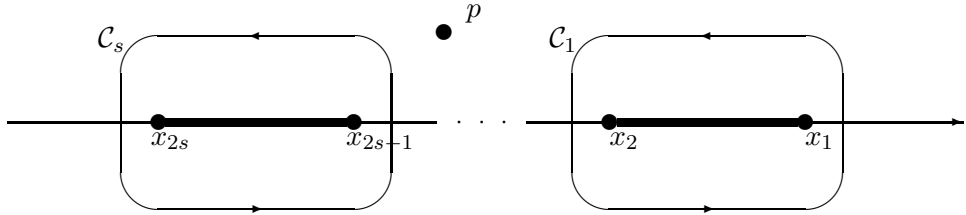

The derivation of the loop equation for multiple cuts can be 
performed exactly along the same lines like for the one-cut case,
exploiting the invariance of the partition function under a 
field redefinition $\phi \rightarrow \phi + \epsilon/(p-\phi)$.
The result differs only by the contour ${\cal C}$ of the complex integral
(s. fig. \ref{fig1})
\beq
\cI \frac{ V^{\prime}(\om)}{p-\om} W(\om) \ = \ 
   (W(p))^2 + \frac{1}{N^2}\dV(p)W(p) \ ,\ \ p\not\in \sigma \ ,\label{loop}
\eeq
where $V^{\prime}(\om)=\sum_j g_j\om^{j-1}$. In the derivation it has been 
assumed like in the one-cut case (e.g. in \cite{AMBLH94}), 
that the density 
$\rho_N(\lam)\equiv\frac{1}{N}\langle\sum_{i}^N\delta(\lam-\lam_i)\rangle $
of the eigenvalues of the matrices $\phi$ has a compact support $\sigma$
in the vicinity of $N=\infty$. Here $\sigma$ consists of an arbitrary
but fixed number $s$ of distinct intervals
\beq
\sigma\ \equiv\ \bigcup_{i=1}^s [x_{2i},x_{2i-1}] \ ,
\ \ x_1> x_2> \ldots > x_{2s} \ \ .\  
\eeq
From rewriting $W(p)$ in terms of $\rho_N(\lam)$
\beq
W(p) \ =\ \int d\lam \frac{\rho_N(\lam)}{p-\lam} \ , \label{Wrho}
\eeq
it is clear, that when expanded at $N=\infty$ $W(p)$ has $s$ cuts
along the real axis on $\sigma$ and is analytic elsewhere. Hence the 
contour ${\cal C}$ in the loop 
equation (\ref{loop}) has to enclose all singularities of $W(\om)$, 
but not the point $\om =p$. It may now well be situated
between two cuts, in particular a double-scaling limit can be
performed at an internal edge of a cut (s. chap. \ref{dslim}).

Inserting the genus expansion eq. (\ref{Wg}) into the loop equation
(\ref{loop}) one gets 
\beq
\cI \frac{V^{\prime}(\om)}{p-\om} W_0(\om) \ = \ (W_0(p))^2 \label{plan} 
\eeq
for genus zero and
\beq
(\hat{\cal K}-2W_0(p))W_g(p) \ = \ \sum_{g\prime=1}^{g-1}W_{g\prime}(p)
          W_{g-g\prime}(p)+\dV(p)W_{g-1}(p) \ , \ \ g\ge 1 \ ,\label{loopg}
\eeq
for higher genera, where $\hat{\cal K}$ is a linear integral operator
defined by
\beq
\hat{\cal K} f(p) \ \equiv \ \cI \frac{\Vp(\om)}{p-\om}f(\om) \ .
\eeq
Knowing the result for $W_0(p)$ one can determine $W_g(p)$ for $g\geq 1$
iteratively from contributions of lower genus on the r.h.s. of
eq. (\ref{loopg}), provided that the operator $(\hat{\cal K}-2W_0(p))$
can be inverted uniquely.

\sect{The planar solution}\label{planar}

\indent

The quadratic equation (\ref{plan}) for the planar solution $W_0(p)$
can be solved in the following way. Deforming the contour in eq.
(\ref{plan}) to infinity one gets
\beq
(W_0(p))^2 \ = \ \Vp(p) W_0(p)
 +\oint_{\cal C_{\infty}}\frac{d\om}{2\pi i} \frac{\Vp(\om)}{p-\om}W_0(\om) \ .
\eeq
The solution is then formally given by
\beql
W_0(p) &=& \frac{1}{2}\Vp(p) - \frac{1}{2}\sqrt{(\Vp(p))^2+4Q(p)} \ , 
\nonumber\\
Q(p)   &=& \oint_{\cal C_{\infty}}\frac{d\om}{2\pi i}
      \frac{\Vp(\om)}{p-\om}W_0(\om) \ ,
\eeql
where the minus sign is chosen in order to fulfill the asymptotic
eq. (\ref{Winf}).
Now if it is assumed that $W_0(p)$ has $s$ cuts in the complex plane,
one can make the following ansatz for the square root,
\beq
W_0(p) \ = \ \frac{1}{2}\lp \Vp(p)-M(p)
            \sqrt{\prod\nolimits_{i=1}^{2s}(p-x_i)}\rp \ . \label{ansatz} 
\eeq
$M(p)$ is assumed to be an $analytic$ function, which still has to be 
determined. The signs of the complex square root are defined by the following
choice,
\beq
\lim_{p\to\pm\infty}\ \sqrt{\prod\nolimits_{i=1}^{2s}(p-x_i)} \ \sim \ 
  p^{s}\ .
\eeq
The analyticity of $M(p)$ allows one to write it as
\beq
M(p) \ = \ \oint_{\cal C_{\infty}}\frac{d\om}{2\pi i}\frac{M(\om)}{\om-p} \ .
\label{Ma}
\eeq
Now solving eq. (\ref{ansatz}) for $M(\om)$ and plugging it into eq. 
(\ref{Ma}) yields
\beq
M(p) \ = \ \oint_{\cal C_{\infty}} \frac{d\om}{2\pi i}
\frac{\Vp(\om)}{(\om-p)\sqrt{\prod_{i=1}^{2s}(\om-x_i)}}\label{M} \ ,
\eeq
because the $W_0(\om)$-term vanishes due to its asymptotic behaviour
at infinity $\sim \frac{1}{\om}$. When evaluated for a given potential of
finite degree $d$, $M(p)$ will be a polynomial of degree $d-s-1$.
Therefore the maximal possible number of cuts has to be 
$s_{max}\leq d-1$ \footnote{It can be shown, that even 
$s_{max}\leq \frac{d}{2}$ holds.}.
Reinserting the solution for $M(p)$ into eq. (\ref{ansatz}) and deforming
back the contour leads to the final result, the planar one-loop correlator
with an $s$-cut structure,
\beq
W_0(p) \ = \ \frac{1}{2}\cI \frac{\Vp(\om)}{p-\om}
           \sqrt{\prod_{i=1}^{2s}\lp \frac{p-x_i}{\om-x_i}\rp} \ .\label{W0}
\eeq
From this result the planar eigenvalue density $\rho(\lam)\equiv
\lim_{N\to\infty}\rho_N(\lam)$ can be recovered, taking the limit $N\to\infty$
of eq. (\ref{Wrho})
\beql
\rho(\lam)\ &=& \ \frac{1}{2\pi i}\lim_{\epsilon\to 0}
         \Big( W_0(\lam-i\epsilon)-W_0(\lam+i\epsilon) \Big) \nonumber\\ 
          \ &=& \ \frac{1}{2\pi}|M(\lam)|
                  \sqrt{-\prod\nolimits_{i=1}^{2s}(\lam-x_i)}
                 \ ,\ \ \ \ \ \ \lam \in \sigma \ . \label{rho}
\eeql
Equivalently $\rho(\lam)$ could have been obtained from the saddle point
equation for the partition function in terms of the eigenvalues
(see for example in \cite{LE91}).
Here the solution takes a closed form for an arbitrary number of cuts $s$.

In the planar solution for $W_0(p)$ eq. (\ref{W0}) the edges of the cuts
$x_i$, $i=1,\ldots,2s$ , still have to be determined as functions of 
the coupling constants $g_i$, $i\in N_+$. These boundary conditions are 
derived by exploiting again the asymptotic behaviour of $W_0(p)\sim 
\frac{1}{p}$. All the coefficients in eq. (\ref{W0}) of the order
$p^{s-1}$ down to $p^0$ must vanish, whereas the $\frac{1}{p}$-coefficient
must equal unity,
\beq
\delta_{k,s} \ = \ \frac{1}{2} 
 \cI \frac{\om^k V^{\prime}(\om)}{\sqrt{\prod_{i=1}^{2s}(\om-x_i)}}
           \ ,\ \ k=0,\ldots,s \ .\label{Rand1}
\eeq
These conditions provide only $s+1$ equations for the $2s$ constants $x_i$.
This is sufficient if and only if $s=1$. For $s\geq 2$ the remaining equations
are given by a criterion of stability against tunnelling of eigenvalues
between different cuts in the limit $N\to\infty$ \cite{DAV90}.
Stability is achieved by imposing equality of the chemical potentials of all 
cuts in the saddle point equation for the density $\rho(\lam)$.
These conditions can be most suitably written in the form \cite{JU90}
\beq
0\ = \ \int\limits_{x_{2k+1}}^{x_{2k}}d\lam\ M(\lam)
           \sqrt{\prod\nolimits_{i=1}^{2s}(\lam-x_i)}\ ,
\ \ \ \ \ \ k=1,\ldots,s-1\ ,
\label{Rand2}
\eeq
where the integration is performed between each two neighbouring cuts. 
This second type of boundary condition for $s\geq 2$ leads to the 
appearance of elliptic integrals in the higher genus correlators as well
as in the planar multi-loop correlators.

\sect{The iterative procedure}\label{iterate}

\indent

After having solved the planar part of $W(p)$ as a starting point,
higher genus contributions can now be iteratively determined by inverting 
the genus expanded loop equation (\ref{loopg}). From $W_g(p)$ all
multi-loop correlators of the same genus can be obtained then by simply
applying the loop insertion operator $\dV \scr (p)$ to it (see eq. 
(\ref{Wdef}) and (\ref{W})).
Along the same lines like in the one-cut solution \cite{AMB93}
a change of variables from coupling constants to moments $M_i^{(k)}$
will allow to express the higher genus correlators nonperturbatively
in the coupling constants $g_i$, depending then only on a finite number of 
these moments,
\beq
M_i^{(k)}\ \equiv \ \cI \Vp(\om)\phi_i^{(k)}(\om) \ ,\ \ \ \ k\in \mbox{N}_+,
\ \ i=1,\ldots,2s \ ,\label{Momente}
\eeq
with
\beq
\phi_i^{(k)}(\om) \ \equiv\ \frac{1}{(\om-x_i)^k} \ \pho(\om) \ ,\ \ \ \ 
\pho(\om)         \ \equiv\ \frac{1}{\sqrt{\prod_{i=1}^{2s}(\om-x_i)}} \ .
\eeq
Deforming the contour to infinity and expanding $\phi_i^{(k)}(\om)$
the moments can be seen to depend in the following way on the couplings,
\beq
M_i^{(k)} \ =\ g_{k+s} +  g_{k+s+1} \sum_{j=1}^{2s}
               x_j(\frac{1}{2} +k\delta_{ij}) +\ldots \ .\label{Mgi}
\eeq
Here an explicit dependence on the number of cuts $s$ enters. The moments
may also be used to characterise multi-critical points. Namely because of
\beq
M_i^{(k)} \ =\ \frac{1}{(k-1)!} \ \frac{d^{k-1}}{d\lam^{k-1}}
               M(\lam)\Big |_{\lam=x_i}   \ \ ,       \label{Mtayl} 
\eeq
the $m$-th multi-critical point at $x_i$ is given by
\beql
M_i^{(k)} \ &=& 0 \ ,\ \ k=1,\ldots ,m-1 \ , \nonumber\\
M_i^{(m)} &\not=& 0 \ ,
\eeql
which indicates that the eigenvalue density $\rho(\lam)$ in eq. (\ref{rho})
accumulates $m-1$ extra zeros at $x_i$.

Secondly it can be seen from eqs. (\ref{Mtayl}) and (\ref{ansatz}), that the 
planar one-loop correlator depends on all moments, when $M(\lam)$ is expanded 
in a Taylor series. This is a special feature of genus $g=0$ only.

\subsection{Determination of the basis}

\indent

In order to proceed it is necessary to introduce a basis for the operator
acting on $W_g(p)$ in eq. (\ref{loopg}),
\beq
(\hat{\cal K}-2W_0(p))\ \chi_i^{(n)}(p) \ \equiv \  \frac{1}{(p-x_i)^n} \ , 
\ \ \ \ n\in \mbox{N}_+ , \ i=1,\ldots,2s  \ .\label{Basis}
\eeq
Given that the r.h.s. of eq. (\ref{loopg}) is a fractional rational 
function of $p$ having poles at the $x_i$ only, $W_g(p)$ will then 
have the following structure
\beq
W_g(p) \ =\ \sum_{n=1}^{3g-1}\sum_{i=1}^{2s} A_{i,g}^{(n)} \chi_i^{(n)}(p)\ ,
\ \ g\ge 1\ . \label{Wstr}
\eeq
The $A_{i,g}^{(n)}$ are complicated functions of the $x_i$ and the 
moments $M_i^{(k)}$. As the order of the highest pole in $W_g(p)$
is not changed by assuming a multi-cut structure, $W_g(p)$ will 
depend on at most $2s(3g-1)$ moments, following the same arguments like in the 
one-cut solution \cite{AMB93}.

A set of basis functions fulfilling eq. (\ref{Basis}) is defined by
\beq
\tilde{\chi}_i^{(n)}(p) \equiv \frac{1}{M_i^{(1)}}\lp \phi_i^{(n)}(p)-
                   \sum_{k=1}^{n-1}M_i^{(n-k+1)}\tilde{\chi}_i^{(k)}(p) \rp 
              \ , \ \ n\in \mbox{N}_+ \ , \ i=1,\ldots,2s \ \ , \label{chi}
\eeq
which can be proven by induction. However, this definition is not unique,
as the kernel of $(\hat{\cal K}-2W_0(p))$ is not empty,
\beq
\mbox{Ker}(\hat{\cal K}-2W_0(p)) \ =\ \mbox{Span}\{p^l\pho(p);\ 
                                 l=0,\ldots, s \} \ .
\eeq
This can be shown by using the boundary conditions of the first type eq.
(\ref{Rand1}). Now because of the asymptotic behaviour 
eq. (\ref{Winf}) only terms $\sim {\cal O}(\frac{1}{p^2})$ at large $p$ may be
added to $W_g(p)$ for $g\geq 1$. This requirement reduces the set of 
zero modes to be
\beq
\mbox{allowed zero modes}: \ \ \{  p^l\pho(p); \ l=0,\ldots,s-2 \} \ .
\label{Nm}
\eeq
In particular for the one-cut solution this argument excluded
any zero mode to be added, making the definition of the basis in eq.
(\ref{chi}) unique. In the general case any linear combination of the $s-1$ 
functions can in principle be added to $W_g(p)$ for a solution with $s$ cuts.

In addition to its asymptotic $W_g(p)$ has to fulfill eq. (\ref{dWg}),
which can be used to fix the basis uniquely. It says that the $p$-dependence
of $W_g(p)$ must be completely absorbable into derivatives with respect to
$\dV\scr (p)$. This will only be possible, if the basis functions 
$\chi_i^{(n)}(p)$ may be expressed completely in terms of 
$\frac{dx_i}{dV}\scr (p)$ and $\frac{dM_i^{(k)}}{dV}\scr (p)$ as 
functions of $p$. So the $\tilde{\chi}_i^{(n)}(p)$ in eq. (\ref{chi})
must be redefined in order to achieve this, which will precisely fix
its zero mode content. The derivatives of the $x_i$ and the moments
$M_i^{(k)}$ can be obtained by applying the loop insertion operator
$\dV\scr (p)$ to the definitions (\ref{Momente}) 
and to the boundary conditions, when rewriting it in the following way:
\beql
\dV (p) \ &=&\ \parV (p) + \sum_{i=1}^{2s}\dxi \frac{\partial}{\partial x_i} 
  \ + \ \sum_{i=1}^{2s}\sum_{k=1}^{\infty}\frac{dM_i^{(k)}}{dV}(p) 
          \frac{\partial}{\partial M_i^{(k)}} \ ,\nonumber\\
\parV (p) \ &\equiv & \ - \sum_{j=1}^\infty \frac{j}{p^{j+1}} 
                      \frac{\partial}{\partial g_j} \ . \label{dVsum}
\eeql
Using the identity
\beq
\parV (p) \Vp(\om) \ =\ \frac{-1}{(p-\om)^2} \ \ ,
\label{pardV}
\eeq
the result for the moments reads
\beql
\frac{dM_i^{(k)}}{dV}(p)&=& 
     (k+\frac{1}{2}) \lp M_i^{(k+1)} \dxi-\phi_i^{(k+1)}(p) \rp\nonumber\\
  &&+\ \frac{1}{2} \sum_{j=1 \atop j\not= i}^{2s}
      \sum_{l=1}^k \frac{1}{(x_j-x_i)^{k-l+1}} \Big( \phi_i^{(l)}(p)-
          M_i^{(l)} \frac{dx_j}{dV}(p) \Big)                  \nonumber\\
  &&+\ \frac{1}{2} \sum_{j=1 \atop j\not= i}^{2s}  \frac{1}{(x_j-x_i)^k}
           \Big( M_j^{(1)}\frac{dx_j}{dV}(p)-\phi_j^{(1)}(p) \Big)  
      \ , \nonumber\\
   &&\ \ i=1,\ldots,2s \ , \ k\in\mbox{N}_+ \ .   
                                                            \label{dM}
\eeql
The quantities $\frac{dx_i}{dV}\scr (p)$ 
are given by the solution of the following set of linear equations,
\beql
0&=& \sum_{i=1}^{2s}\lp x_i^k M_i^{(1)}\dxi \ - \ p^k\phi_i^{(1)}(p)\rp
     \ + \ 2k\ p^{k-1}\pho(p) \ , \ \ k=0,\ldots,s \ , \nonumber\\
0&=& \sum_{i=1}^{2s}\lp M_i^{(1)}\dxi\ - \ \phi_i^{(1)}(p)\rp K_{i,j} \ , 
    \ \ \ \ \ \ \ \ \ \ j=1,\ldots,s-1\ ,  \nonumber\\
&&K_{i,j} \ \equiv \int\limits_{x_{2j+1}}^{x_{2j}}d\lam
    \frac{\sqrt{\prod_{k=1}^{2s}(\lam-x_k)}}{(\lam-x_i)}\ ,
 \label{dx}
\eeql
where $\dV\scr (p)$ has been applied to eqs. (\ref{Rand1}) and (\ref{Rand2}).
The result of the latter is derived in appendix \ref{A}, where some care
has to be taken to the interchanging of $\dV\scr (p)$ and deforming the 
contour in eq. (\ref{M}) for $M(\lam)$ to infinity.
It can easily be seen from the linear system of equations (\ref{dx}), that the
solution will always take the form
\beq
M_i^{(1)}\dxi \ = \ \phi_i^{(1)}(p)\ +\ \sum_{l=0}^{s-2}\al_{i,l}\ p^l\pho (p) 
\ \ , \ \ \ \ \ i=1,\ldots,2s  \ \ .\label{dxsol}
\eeq
The $\al_{i,l}$, $i=1,\ldots,2s$, $l=0,\ldots,s-2$, will then be determined 
by the following set of equations:
\beql
0&=& \sum_{i=1}^{2s}\sum_{l=0}^{s-2}\al_{i,l}\ x_i^k\ p^l \ -\ 
     \sum_{i=1}^{2s}\sum_{l=0}^{k-1}x_i^{k-1-l}\ p^l \ +\ 2k\ p^{k-1}\ ,
     \ \ k=0,\ldots,s \ , \nonumber\\
0&=& \sum_{i=1}^{2s}\sum_{l=0}^{s-2}\al_{i,l}\ K_{i,j}\ p^l  \ ,
     \ \ j=1,\ldots,s-1 \ . \label{alsyst}
\eeql
Comparing the coefficients in powers\footnote{The coefficient of $p^{k-1}$
is zero for $k=s$.} of $p^l$, $l=0,\ldots,,s-2$, leads to $2s(s-1)$ equations
for the $\al_{i,l}$. They will only depend on the $x_i$ and 
$K_{i,j}$. Hence the only difference between the $\frac{dx_i}{dV}\scr (p)$
in eq. (\ref{dxsol}) and the respective quantities in the one-cut solution 
\cite{AMB93} is the occurrence of a linear combination of the allowed
zero modes from eq. (\ref{Nm}). More explicit results for $s=2$ are given in 
the next chapter.

Now all necessary ingredients for redefining the basis (\ref{chi}) in terms of
total derivatives are collected. Solving eq. (\ref{dM}) for 
$\phi_i^{(k+1)}(p)$ and expressing its $p$-dependence as $\dV\scr (p)$-terms 
plus zero modes,
a unique basis can now be obtained inductively from eq. (\ref{chi}) by 
subtracting these corresponding zero modes:
\beql
\chi_i^{(n)}(p) &\equiv& \frac{1}{M_i^{(1)}}\lp 
                \phi_i^{(n)}(p)\Big |_{\dV -part}-
         \sum_{k=1}^{n-1}M_i^{(n-k+1)}\chi_i^{(k)}(p) \rp \ , \nonumber\\
         &&\ \ i=1,\ldots,2s \ ,\ n\in \mbox{N}_+ \ \ . \label{chitot}
\eeql
The first basis functions then read
\beql
\chi_i^{(1)}(p) &=& \dxi \ ,\ \ \ \ \ \ \ \ i=1,\ldots,2s\ , \nonumber\\
\chi_i^{(2)}(p) &=& -\frac{2}{3}\dV (p) \ln |M_i^{(1)}| -\frac{1}{3}
          \sum^{2s}_{j=1 \atop j\neq i} \dV(p) \ln |x_i-x_j| \ \ .
\label{basis2}
\eeql

\subsection{Calculation of genus 1}

\indent

Having determined the basis, the loop equation can now be inverted step
by step in genus. For genus $g=1$ eq. (\ref{loopg}) reads
\beq
(\hat{\cal K}-2W_0(p))W_1(p) \ = \ \dV(p)W_0(p) \ . \label{loop1} 
\eeq
Using the result for $W_0(p)$ eq. (\ref{W0}) and the loop insertion
operator from eq. (\ref{dVsum}) the r.h.s. is given by
\beql
\dV(p)W_0(p)    &=& -\frac{3}{16}\sum_{i=1}^{2s}\frac{1}{(p-x_i)^2}     
           \ -\ \frac{1}{8}\sum_{i,j=1 \atop i<j}^{2s}\frac{1}{(p-x_i)(p-x_j)}
             \nonumber\\
           &&+\ \frac{1}{4}\frac{1}{\pho(p)}\sum_{i=1}^{2s}\frac{1}{p-x_i}
               M_i^{(1)} \dxi \nonumber\\
             &=&\ \frac{1}{16}\sum_{i=1}^{2s}\frac{1}{(p-x_i)^2}     
                 \ -\ \frac{1}{8}\sum_{i,j=1 \atop i<j}^{2s}
                 \frac{1}{x_i-x_j}\lp\frac{1}{(p-x_i)}-\frac{1}{(p-x_j)}\rp
              \nonumber\\
           &&+\ \frac{1}{4}\sum_{i=1}^{2s}
                 \sum_{l=0}^{s-2}\frac{\al_{i,l}\ x_i^l}{p-x_i}   \ \ .
 \label{dW0}
\eeql
Here the fact has been used, that all the regular parts coming from
$\frac{p^l}{p-x_i}$, $l=1,\ldots,s-2$, will  vanish due to eq. (\ref{alsyst}),
such that $W_0(p,p)=\dV{\scr (p)}W_0(p)$ fulfils its correct asymptotic.
The two-loop correlator at different arguments $W_0(p,q)$ can be obtained
in the same way, which is derived in appendix \ref{B}.
The result for the one-loop correlator of genus one with $s$ cuts can now
easily be obtained by using the basis eq. (\ref{basis2}),
\beql
W_1(p)  &=& \frac{1}{16}\sum_{i=1}^{2s}\chi_i^{(2)}(p)
                 \ -\ \frac{1}{8}\sum_{i,j=1 \atop i<j}^{2s}
           \frac{1}{x_i-x_j}\lp \chi_i^{(1)}(p)-\chi_j^{(1)}(p)\rp \nonumber\\
            && + \frac{1}{4}\sum_{i=1}^{2s}
                 \sum_{l=0}^{s-2}\al_{i,l}\ x_i^l\ \chi_i^{(1)}(p)   \ \ .
\label{W1}
\eeql
The integration of $W_1(p)$ in order to get $F_1$ as well as the calculation
of higher genera gets technically very much involved, as the integrals $K_{i,j}$
then have to be integrated or differentiated with respect to the $x_i$, 
$i=1,\ldots,2s$.

For the case of two cuts, the integrals $K_{i,1}$, $i=1,\ldots,4$ , can be
expressed by the well known complete elliptic integrals of the first, second
and third kind, making a more detailed analysis possible. This will be
the subject of the next chapter.

\sect{The two-cut solution}\label{twocut}

\indent

The case of two cuts may already appear for the symmetric quartic potential,
when the coupling constants are suitably chosen \cite{CMM86}. The approach
presented here allows for a closed treatment of an arbitrary polynomial 
potential including higher genus contributions. It is beyond the scope of
the method of orthogonal polynomials, as it has been mentioned already in the
introduction.

The explicit solution for the zero mode coefficients 
$\al_{i,0}$ in eq. (\ref{dxsol}) will allow to inspect
the planar two-loop correlator in more detail. Knowing the complete
$x_i$-dependence of these coefficients makes it also possible to study 
the double-scaling limit.
The set of equations (\ref{alsyst}) reads for $s=2$
\beql
0 &=& \sum_{i=1}^4 \al_i \nonumber\\
0 &=& \sum_{i=1}^4 \al_i x_i\ \ -\ 2 \nonumber\\
0 &=& \sum_{i=1}^4 \lp \al_i x_i^2 -x_i\rp \nonumber\\
0 &=& \sum_{i=1}^4 \al_i  K_i  \ ,\label{dx2}
\eeql
where $\al_i \equiv \al_{i,0}$ and
\beq
K_i \ \equiv\ \bI d\lam 
   \frac{\sqrt{\prod_{j=1}^4 (\lam-x_j)}}{(\lam-x_i)} \ ,\ \ i=1,\ldots,4 \ .
   \label{Ki}
\eeq
The solution for the integrals $K_i$ in terms of complete elliptic
integrals may be taken from \cite{Byrd},
the result being displayed in appendix \ref{C}.
The solution of the eqs. (\ref{dx2}) reads
\beq
M_i^{(1)}\dxi \ =\ \phi_i^{(1)}(p) \ +\ \al_i\ \pho(p) 
          \ , \ \ i=1,\ldots,4 \ ,\label{dxi2}
\eeq
where the $\al_i$ can be found after a tedious calculation to be
\beql
\al_1 &=& \frac{1}{x_1-x_4}\lp 1+ 
            \frac{(x_2-x_4)}{(x_1-x_2)}\frac{E(k)}{K(k)}\rp \ ,\nonumber\\  
\al_2 &=& \frac{1}{x_2-x_3}\lp 1+ 
            \frac{(x_1-x_3)}{(x_2-x_1)}\frac{E(k)}{K(k)}\rp \ ,\nonumber\\
\al_3 &=& \frac{1}{x_3-x_2}\lp 1+ 
            \frac{(x_2-x_4)}{(x_4-x_3)}\frac{E(k)}{K(k)}\rp \ ,\nonumber\\
\al_4 &=& \frac{1}{x_4-x_1}\lp 1+ 
            \frac{(x_1-x_3)}{(x_3-x_4)}\frac{E(k)}{K(k)}\rp \ ,\nonumber\\ 
&& \nonumber\\ 
&&k^2\ =\ \frac{(x_1-x_4)(x_2-x_3)}{(x_1-x_3)(x_2-x_4)} \ .
\label{alf}
\eeql
Having the explicit result for the $\frac{dx_i}{dV}\scr (p)$ at hand, 
the planar two-loop correlator can be evaluated also for different arguments
most explicitly. It is derived in appendix \ref{B}, which reads
\beql
W_0(p,q)&=&\frac{1}{4(p-q)^2}\Bigg(
  \sqrt{\frac{(p-x_1)(p-x_4)(q-x_2)(q-x_3)}{(p-x_2)(p-x_3)(q-x_1)(q-x_4)}}
\nonumber\\
&&\ \ \ \ \ \ \ \ \ \ \ \ 
   +\ \sqrt{\frac{(p-x_2)(p-x_3)(q-x_1)(q-x_4)}{(p-x_1)(p-x_4)(q-x_2)(q-x_3)}}
 \Bigg)\nonumber\\
      &+&\frac{1}{4}\ \frac{1}{\sqrt{\prod_{j=1}^4(p-x_j)(q-x_j)}}
         \frac{E(k)}{K(k)} f(\{x_i\})
     -\frac{1}{2(p-q)^2} \ .
\label{Wpqs} 
\eeql
The function $f(\{x_i\})$ of the $x_i$ is given in eq. (\ref{W0pq2f}).
The result fulfils its required analyticity properties (see appendix
\ref{B}). As it must be regular at coinciding arguments, it has to be 
compared with eq. (\ref{dW0}) for the two-cut solution, 
which can also be found in appendix \ref{B},
\beql
W_0(p,p)&=& \frac{1}{16}\sum_{i=1}^4\frac{1}{(p-x_i)^2}
   +\frac{1}{4}\lp \frac{1}{(p-x_1)(p-x_4)}+\frac{1}{(p-x_2)(p-x_3)}\rp
  \nonumber\\
         &-& \frac{1}{8}\sum^4_{i<j}\frac{1}{(p-x_i)(p-x_j)}
      +\frac{1}{4}\frac{1}{\prod_{j=1}^4(p-x_j)} \frac{E(k)}{K(k)} 
           f(\{x_i\}) \ .
\label{Wpps}                
\eeql
These results for the planar two-loop correlator are clearly universal,
as they depend on the coupling constants only implicitly via the endpoints
of the cuts. The part proportional to the elliptic integrals simplifies 
considerably in the limit of a symmetric potential, where $x_4=-x_1$ and
$x_3=-x_2$. The modulus will then be
\beq
k^2_{sym}\ =\ \frac{4x_1x_2}{(x_1+x_2)^2} \ ,
\eeq
and the function $f(\{x_i\})$ becomes
\beq
f_{sym}(\{x_i\}) \ = \ (x_1+x_2)^2 \ .
\eeq

\subsection{Results for genus $g=1$}

\indent

With the more explicit results for the two-cut solution at hand
the free energy $F_1$ can now be integrated from $W_1(p)$. 
As it has been mentioned already
the knowledge of $F_1$ in not necessary to get the higher orders of $W_g(p)$
though. Inserting the basis functions eq. (\ref{basis2}) into eq. (\ref{W1})
$W_1(p)$ reads
\beql
W_1(p) &=& -\frac{1}{24}\sum_{i=1}^4\dV (p)\ln |M_i^{(1)}|
              -\frac{1}{6}\sum_{i<j}\dV (p)\ln |x_i-x_j| \nonumber\\
         &&+\frac{1}{4}\sum_{i=1}^4\al_i\dxi \ .
\label{W1p2}
\eeql
Using the well known relation for the integral $K(k)$,
\beq
\frac{\partial}{\partial k^2} K(k) \ =\ \frac{1}{2k^2(1-k^2)}
         \Big( E(k)-(1-k^2)K(k) \Big) \ , \label{dK}
\eeq
the following helpful relation can be derived,
\beq
2\dV (p) \ln |K(k)| \ = \ -\sum_{i=1}^4\al_i\dxi \ + \ 
  \dV (p)\ln |x_1-x_3|  +\dV (p)\ln |x_2-x_4| \ .
\label{dlnK}
\eeq
The free energy $F_1$ now can be read off easily 
from eq. (\ref{dWg}), $W_1(p)=\dV{\scr (p)} F_1$,
\beql
F_1 &=& -\frac{1}{24}\sum_{i=1}^4\ln |M_i^{(1)}| -\frac{1}{2}\ln |K(k)| 
         - \frac{1}{6}\sum_{i<j}\ln |x_i-x_j| \nonumber\\   
    &&+\frac{1}{4}\Big( \ln |x_1-x_3|+\ln |x_2-x_4|\Big) \ .
\eeql
Taking the symmetric limit\footnote{The moments will then behave like
$M_4^{(n)}=(-)^nM_1^{(n)}$, $M_3^{(n)}=(-)^nM_2^{(n)}$.}, 
the result looks very similar to the
free energy of genus one of the $O(n)$-model for $n=+2$ \cite{EK95}.

The calculation of higher genera may now be performed by computer algebra,
using eq. (\ref{loopg}). Taking the derivative $\dV\scr(p)$, all quantities
can expressed algebraically by the known expressions for the 
$\frac{dx_i}{dV}\scr (p)$, $\frac{dM_i^{(k)}}{dV}\scr (p)$ and $K(k)$ 
and $E(k)$, using eq. (\ref{dK}) and a similar expression for $E(k)$.

The particular case of having two cuts allows for a certain check of
the one-loop correlator, which is valid for all genera. If one is taking
the symmetric limit at the end of the calculation, which is implemented 
by setting the odd coupling constants $g_{2i+1}$, $i\in$ N, to zero,
all the odd powers of expectation values of matrices will vanish,
\beq
\langle \mbox{Tr} \phi^{2k+1} \rangle_{sym} \ =\ 0 \, \ \ k\in \mbox{N} \ .
\eeq
Consequently $W(p)$ will become an odd function in $p$,
\beql
W^{sym}(p) &=&\ \frac{1}{N}\left\langle \mbox{Tr}
                   \frac{1}{p-\phi}\right\rangle_{sym}  
             \ =\ \frac{1}{N} \sum_{k=0}^{\infty} 
       \frac{\langle \mbox{Tr} \phi^{2k} \rangle_{sym}}{p^{2k+1}} \\
           &=&\ \frac{1}{p} \ +\ \frac{1}{N}\ 
       \frac{\langle \mbox{Tr} \phi^{2} \rangle_{sym}}{p^{3}} \ +\ \ldots \ .
       \label{Wsym}
\eeql
In particular the term $\sim {\cal O}(\frac{1}{p^2})$ will disappear for all
genera. Now the only function of $p$ contained in the basis with this 
asymptotic behaviour\footnote{For $s\geq 3$ there will always be zero
modes left with an asymptotic $\sim {\cal O}(\frac{1}{p^3})$.}
is the zero mode $\pho (p)$. Hence it is not 
contributing to the symmetric one-loop correlator $W_g^{sym}(p)$ for all
genera. Using this special property the result for $W_1(p)$ can be checked
in this limit. Extracting the $\pho (p)$-dependence from eqs. (\ref{W1p2}) 
and (\ref{dlnK}) it can be shown, that its coefficient indeed vanishes in the 
symmetric limit.

\newpage

\sect{The double-scaling limit of the two-cut solution} \label{dslim}

\indent

It has been argued that in the double-scaling limit (d.s.l.) all the 
multi-cut solutions should be equivalent to the one-cut solution for the 
following reason. Adjusting the coupling constants to achieve a
multi-critical behaviour at one of the edges of the cuts, say at $x_j$,
the d.s.l. magnifies the accumulation of zeros at $x_j$ such that the structure
elsewhere can be neglected. However, this argument is only true, if
the cuts do not touch or vanish at the same time.
The given solution with two cuts therefore 
supplies an example, where this can be checked explicitly. It will be 
verified, that the generic d.s.l. at any $x_j$ is equivalent to the d.s.l. 
of the one-cut solution \cite{AMB93}. 
But if the scaling limit is taken at a cut, which simultaneously 
shrinks to zero, or merges with the other cut, or both, a different 
continuum behaviour will turn out.

\subsection{Scaling limit at $x_j$}

\indent

In the generic case the d.s.l. is performed at one specific edge of
the cuts $x_j$, where for the $m$-th multi-critical point $m-1$ extra
zeros accumulate in the eigenvalue density eq. (\ref{rho}). 
Fixing the coupling constants appropriately to reach
this point, $p$ and $x_j$ will scale in the following way
\beql
x_j &=& x_j^c \ -\ a \Lam^{\frac{1}{m}} \nonumber\\
p \ &=& x_j^c \ +\ a\pi \ , \label{dsl}
\eeql
whereas the $x_{i\neq j}$ are kept fixed. For an $m$-th multi-critical point
the moments scale according to 
\beq
M_j^{(k)}\ \sim\ a^{m-k}\ \ , \ \ k=1,\ldots,m-1 \ ,
\label{Mdsl}
\eeq
where again the $M_{i\neq j}^{(k)}$ do not scale. Looking at the explicit
solution for the first basis function, which is given by eqs. (\ref{dxi2}) and
(\ref{alf}), it is clear, that the zero mode contribution will be 
sub-dominant in the scaling limit eq. (\ref{dsl}),
\beq
\frac{dx_j}{dV}(p)\ =\ \frac{1}{M_j^{(1)}}\phi_j^{(1)}(p) 
\ \ \ \ \mbox{(d.s.l.)} \ .\label{dxjdsl}
\eeq
The derivatives of the moments take the form
\beql
\frac{dM_j^{(k)}}{dV}(p)&=& 
   (k+\frac{1}{2}) \lp M_j^{(k+1)}\frac{dx_j}{dV}(p) -\phi_j^{(k+1)}(p) \rp 
   \ ,\nonumber\\
    \phi_j^{(k+1)}(p)&=& \frac{1}{(p-x_j)^{k+1}}
              \frac{1}{\sqrt{(p-x_j)\prod_{i\neq j}(x_j^c-x_i)}}
\ \ \mbox{(d.s.l.)}\ .\label{dMdsl}
\eeql
Looking at the inductive construction of the basis eq. (\ref{chitot})
it is obvious then, that the zero modes will be also sub-dominant for 
the rest of the basis, which is hence given by
\beql
\chi_j^{(n)}(p) &\equiv& \frac{1}{M_j^{(1)}}\lp \phi_j^{(n)}(p)-
                   \sum_{k=1}^{n-1}M_j^{(n-k+1)}\chi_j^{(k)}(p) \rp 
              \ , \ \ n\in \mbox{N}_+ \ \ \ \ \mbox{(d.s.l.)}\nonumber\\
                   &\sim& \ a^{-m-n+\frac{1}{2}} \ .
\eeql
Consequently in the scaling limit eq. (\ref{dsl}) the basis for the poles at
$x_j$ precisely looks like the one for the one-cut case at $x=x_j$
\cite{AMB93}. The other basis functions, which are sub-dominant 
$\sim a^{\frac{1}{2}}$, will not be needed, as only contributions 
from poles at $x_j$ are dominant. The complete equivalence 
to the one-cut solution in the d.s.l. will now be shown by looking at
the starting point of the iteration $W_0(p,p)$ and at the scaled
loop insertion operator. 

It is easy to see from eq. (\ref{dW0}), that the leading part in orders of 
$a$ will be
\beq
W_0(p,p)\ =\ \frac{1}{16}\frac{1}{(p-x_j)^2} \ \ \ \ \mbox{(d.s.l.)}\ ,
\eeq
which is the same like in the one-cut solution. The scaled $\dV \scr (p)$
from eq. (\ref{dVsum}) will take the form
\beq
\dV (p)\ =\ \frac{dx_j}{dV}(p)\frac{\partial}{\partial x_j}\ +\ 
     \sum_{k=1}^{\infty}\frac{dM_j^{(k)}}{dV}(p)
             \frac{\partial}{\partial M_j^{(k)}} \ \ \ \ \mbox{(d.s.l.)}\ ,
\eeq
where $\frac{dx_j}{dV}\scr (p)$ and $\frac{dM_j^{(k)}}{dV}\scr (p)$ 
are given by eq. (\ref{dxjdsl}) and eq. (\ref{dMdsl}) respectively.
Hence the generic d.s.l. of the two-cut solution precisely maps 
to the d.s.l. of the one-cut solution in \cite{AMB93}, when the quantities
$d_c$ and $M_k$ there are replaced by $d_c\equiv \prod_{i\neq j}(x_j^c-x_i)$
and $M_j^{(k)}$ respectively here. However, one has to bare in mind, that in
this symbolical equivalence of the respective formulas the moments will
explicitly depend on the number of cuts as functions of the couplings
(see eq. (\ref{Mgi})).

In reference \cite{AMB93} explicit results up to and including genus $g=4$
are presented in the d.s.l., which are valid here as well. The analysis
concerning the possible kinds of combinations of moments, 
that can appear in eq. (\ref{Wstr}), also applies here.

\subsection{Merging and shrinking cuts in the d.s.l.}

\indent

A different scaling behaviour may be expected, when the d.s.l.
is performed at a point, where the two cuts merge, or one of them shrinks
to zero, or both happens together. 
When the case of merging is considered, it has to be taken 
into account, that extra zeros of $M(p)$ lying between the cuts will be
picked up in this limit\footnote{From analyticity properties and the positivity
of the eigenvalue density it follows \cite{DE90}, that $M(p)$ will always 
have at least one real zero between two adjacent cuts for arbitrary $s$.}.
This will yet only affect the order of multi-criticality, but not the 
scaling behaviour of the zero modes. 

The outcome of the 
analysis will be, that in the d.s.l. at either merging or shrinking cuts
the zero mode contributions will stay sub-dominant, although becoming
singular in the latter case. Nevertheless the continuum results
will change in all cases, as in eq. (\ref{dW0}) more terms will contribute.
When the shrinking and merging of the cuts is considered simultaneously,
say $x_2,x_4\to x_3$, even the zero modes will be enhanced such that they
contribute in the d.s.l., leading to another different continuum
theory.

First the situation of the two cuts merging is considered 
in the limit $x_2\to x_3$, which can be parametrised by
\beql
x_2 &=& x_3^c + a\nu \ , \ \ \nu >0 \ ,\nonumber\\
x_3 &=& x_3^c - a\mu \ , \ \ \mu >0 \ ,\nonumber\\
p\ &=& x_3^c + a\pi \ ,\label{dslm}
\eeql
as at the critical point $p-x_3$ will also scale. Looking at the 
explicit results eq. (\ref{alf}) for the zero modes, the modulus $k^2$
will be of the order $a$,
\beq
k^2 \ =\ a(\nu +\mu ) \frac{(x_1-x_4)}{(x_1-x_3^c)(x_3^c-x_4)}\ +\ 
{\cal O}(a^2) \ . \label{kdsl}
\eeq
So the elliptic integrals can be expanded in $k^2$, which reads
\beq
\frac{E(k)}{K(k)}\ =\ 1 - \frac{1}{2}k^2\ +\ {\cal O}(a^2) \ . 
\eeq
The scaling behaviour in $\al_2$ and $\al_3$ coming from the factor
$\frac{1}{x_2-x_3}$ will thus be cancelled, leaving them sub-dominant
as in the generic case,
\beql
\al_1 &=& \frac{1}{x_1-x_3^c} \ , \nonumber\\
\al_2 &=& \frac{1}{x_3^c-x_1}\lp 1-\frac{1}{2}\frac{(x_1-x_4)}{(x_3^c-x_4)}\rp
\ ,\nonumber\\
\al_3 &=& \frac{1}{x_3^c-x_4}\lp 1-\frac{1}{2}\frac{(x_1-x_4)}{(x_1-x_3^c)}\rp
\ ,\nonumber\\
\al_4 &=& \frac{1}{x_4-x_3^c} \ .
\eeql
The coefficients $\al_1$ and $\al_4$ are regular in the limit
eq. (\ref{dslm}) anyway. The result for the $\frac{dx_i}{dV}\scr (p)$,
$i=2,3$ , will therefore be the same like in eq. (\ref{dxjdsl}).
Still, the continuum limit is altered, as in eq.
(\ref{dM}) more terms will survive now in comparison to eq. (\ref{dMdsl})
in the generic case,
\beql
\frac{dM_2^{(k)}}{dV}(p) &=& 
     (k+\frac{1}{2}) \lp \frac{M_2^{(k+1)}}{M_2^{(1)}}\phi_2^{(1)}(p) 
           -\phi_2^{(k+1)}(p) \rp \ ,\nonumber\\
&+& \frac{1}{2}\sum_{l=1}^k\frac{1}{(x_3-x_2)^{k-l+1}}
   \lp\phi_2^{(l)}(p)-\frac{M_2^{(l)}}{M_3^{(1)}}\phi_3^{(1)}(p)\rp \ ,
\label{dMmerge}
\eeql
and the same for the indices 3 and 2 interchanged. The d.s.l.
performed at the interior of the two cuts merging therefore leads to
a new continuum limit of the theory. The starting point for the iterative
solution in the scaling limit reads
\beql
W_0(p,p)\ =\ \frac{1}{16}\sum_{i=2,3} \frac{1}{(p-x_i)^2}
        \ -\ \frac{1}{8}\frac{1}{x_2-x_3}\lp\frac{1}{p-x_2}-\frac{1}{p-x_3}\rp
\ . \label{W0dslm}
\eeql
Using eqs. (\ref{dxjdsl}) and (\ref{dMmerge}) to obtain the double-scaled 
basis, the genus one result can easily be achieved.

Second, the case of the scaling limit is considered, which is performed 
at a cut simultaneously shrinking to zero, say $x_4\to x_3$. It is conveniently
parametrised by
\beql
x_3 &=& x_3^c + a\nu \ , \ \ \nu >0 \ ,\nonumber\\
x_4 &=& x_3^c - a\mu \ , \ \ \mu >0 \ ,\nonumber\\
p\ &=& x_3^c + a\pi \ ,\label{dsls}
\eeql
where the other possible case $x_2\to x_1$
can be obtained by interchanging the 
indices $4\leftrightarrow 1$ and $3\leftrightarrow 2$ everywhere.
The modulus $k^2$ will reach unity in the limit (\ref{dsls}), 
where the integral 
of the first kind $K(k)$ becomes singular. The integrals in eq.
(\ref{alf}) can now be expanded in terms of the complementary modulus
${k^{\prime}}^2$,
\beq
{k^{\prime}}^2 \equiv 1-k^2=a(\nu+\mu)\lp\frac{1}{x_2-x_3^c}
-\frac{1}{x_1-x_3^c}\rp +{\cal O}(a^2) \ . 
\eeq
Consequently the factor $\frac{E(k)}{K(k)}$ vanishes logarithmically,
\beq
\frac{E(k)}{K(k)}\ \sim \ \frac{1}{\ln a}+{\cal O}(a\ln a) \ .
\eeq
This again spoils the possible dominance of the zero modes, the result reading
\beql
\al_1 &=& \frac{1}{x_1-x_3^c} \ , \nonumber\\
\al_2 &=& \frac{1}{x_2-x_3^c} \ , \nonumber\\
\al_3,\al_4 &\sim& \frac{1}{a\ln a} \ .
\eeql
So although becoming singular in the limit (\ref{dsls}) the zero mode 
contributions will still be sub-dominant in the basis eq. (\ref{dxi2}).
The consequences for the iterative procedure in the d.s.l. are the same
like for the merging cuts, where in eqs. (\ref{dMmerge}) and 
(\ref{W0dslm}) the index 2 has to be replaced by 4.

An even different kind of continuum limit can be found, when the process
of merging and shrinking is put together. In this limit the zero mode
contributions in the basis will no longer be sub-dominant. Starting with
the following parametrisation,
\beql
x_2 &=& x_3^c +a\nu \ ,\ \ \nu >0 \ ,\nonumber\\
x_3 &=& x_3^c \nonumber\\
x_4 &=& x_3^c -a\mu \ , \ \ \mu >0 \ ,\nonumber\\
p\ &=&  x_3^c +a \pi \ , \label{dslms}
\eeql
the modulus and also the ratio $\frac{E(k)}{K(k)}$ will stay finite,
\beq
k^2 \ =\ \frac{\nu}{\nu +\mu} + {\cal O}(a) \ .
\eeq
The modulus may thus reach any value in (0,1). 
For the special choice $\nu=\mu$, it holds that
$k^2={k^{\prime}}^2=\frac{1}{2}$. The 
elliptic integrals can then be expressed by the gamma function.
Using Legendre's relation, which then reads 
\beq
2E(k)K(k)- K(k)^2 \ =\ \frac{\pi}{2} \ , 
\eeq
and the special value $K({\scr \frac{1}{\sqrt{2}}}) 
=\frac{1}{4\sqrt{\pi}}(\Gamma ({\scr \frac{1}{4}}))^2$\ , 
the ratio of the integrals is given by
\beq
\frac{E({\scr \frac{1}{\sqrt{2}}})}{K({\scr \frac{1}{\sqrt{2}}})} 
\ =\ \frac{1}{2}+4\pi^2
\frac{1}{\lp\Gamma ({\scr \frac{1}{4}})\rp^4} \ .
\eeq
The $\al_i$, $i=2,3,4$ , will therefore scale in the desired way,
\beql
\al_1 &=& \frac{1}{x_1-x_3^c} \ , \nonumber\\
\al_2 &=& \frac{1}{a\nu}\lp 1-\frac{E(k)}{K(k)}\rp \ , \nonumber\\
\al_3 &=& \frac{1}{a\nu}\lp -1+\frac{\nu +\mu}{\mu}\frac{E(k)}{K(k)}\rp 
\ , \nonumber\\
\al_4 &=& -\frac{1}{a\mu}\frac{E(k)}{K(k)} \ , \ \ \ \ \ \ \ \ \ 
k^2 \ =\ \frac{\nu}{\nu + \mu} \ ,
\eeql
leaving the form of eq. (\ref{dxi2}) unchanged for $i=2,3,4$. 
In this kind of d.s.l. the iteration merely simplifies, in contrast to the 
generic case. The only terms to be left out in eq. (\ref{dW0})
are those with an index 1. 

This ends the short survey of all possible continuum limits to be performed
in the two-cut solution.

\sect{Conclusion and outlook}

\indent

The results presented here complete the perturbative analysis in $\frac{1}{N}$
of the hermitian one-matrix model. All possible solutions classified by the 
number of cuts are derived in an iterative scheme for higher genus 
contributions to the loop correlators, which
generalises the results of \cite {AMB93} for the one-cut solution.
Explicit expressions were given for the one-loop correlator of genus zero and
one and for the planar two-loop correlator at different arguments.
The latter is universal for each number of cuts $s$, providing a 
whole set of new universality classes.

The two-cut solution was displayed in full detail, including the genus one
contribution to the free energy. When investigating the double-scaling limit,
the two-cut solution was proven to be equivalent to the one-cut solution in
the generic case. However, when taking the continuum limit at merging or
shrinking cuts a different behaviour was revealed. A candidate for this new
continuum behaviour is still missing.

Up to now in most investigations of matrix models there has been assumed 
a one-cut structure of the corresponding quantities. The results given here
should make it possible to deal with multi-cut structures also in other
models, where the loop equation techniques have been applied successfully.
Just to name a few examples like 
the O(n)-model \cite{EK95} or the supereigenvalue model \cite{JP95},
particularly in the complex matrix model \cite{AKM92} most explicit 
results should be accessible, completing the analysis of \cite{AKE95}.
The hope is beside finding new universality classes and continuum
limits there as well the 
appearance of new types of critical behaviour, which might still be
hidden in the full space of solutions.

\indent

\begin{flushleft}

\underline{Acknowledgements}: I would like to thank the Niels Bohr
Institute for financial support and its 
warm hospitality, where part of this work was being
done. In particular I wish to thank J. Ambj{\o}rn, C. Kristjansen and
Yu. Makeenko for many valuable hints and discussions.

\end{flushleft}

\newpage

\begin{appendix}

\sect{$\dV\scr (p)$ of the boundary conditions} \label{A}

\indent

While taking the derivative $\dV\scr (p)$ of the first type of boundary
conditions eq. (\ref{Rand1}) is straight forward, the derivative of the 
second type eq. (\ref{Rand2}) 
\beq
0\ = \ \int\limits_{x_{2j+1}}^{x_{2j}}d\lam\ M(\lam)
           \sqrt{\prod\nolimits_{i=1}^{2s}(\lam-x_i)}\ ,\ \ j=1,\ldots,s-1\ ,
\label{RandA}
\eeq
is more sophisticated. When applying $\dV\scr (p)$ to $M(\lam)$ in the 
integrand, the resulting function will no longer be analytic, having an extra 
pole at $p$. Therefore two things have to be taken into account. First,
the contour in the expression for $M(\lam)$ eq. (\ref{M}) cannot be deformed
to infinity but in the end of the calculation\footnote{Otherwise the extra
pole at $p$ would be overlooked.}, taking it to enclose the poles of the 
integrand only,
\beq
M(\lam) \ = \ \oint_{{\cal C}\cup\cal C_{\lam}} \frac{d\om}{2\pi i}
\frac{\Vp(\om)}{(\om-\lam)\sqrt{\prod_{i=1}^{2s}(\om-x_i)}}\label{MA} \ .
\eeq
Second, when applying $\dV\scr(p)$ to eq. (\ref{RandA}) for fixed $j$, the 
argument $p$ must be excluded to be
in the interval of integration, $p\notin [x_{2j+1},x_{2j}]$. Otherwise the
real integral between the cuts does not exist, due to eq. (\ref{pardV}). 
The final result will
have no poles between any of the cuts, allowing for an analytic
continuation of $\frac{dx_i}{dV}\scr (p)$ to $p$ also being between the cuts.

Taking this into account the derivative $\dV\scr (p)$ in the form of 
eq. (\ref{dVsum}) to eq. (\ref{RandA}) yields
\beql
0 &=& \int\limits_{x_{2j+1}}^{x_{2j}}d\lam 
  \oint_{{\cal C}\cup\cal C_{\lam}} \frac{d\om}{2\pi i}\Bigg(
    \frac{1}{(p-\lam)^2(\om-\lam)} \nonumber\\
&& \ \ \ \ \ \ \ \ \ \ \ \ \ \ \ \ \ \ \ + \ \frac{1}{2}\sum_{i=1}^{2s}\dxi  
    \frac{\Vp (\om)}{\om-\lam}\lp \frac{1}{\om-x_i}-\frac{1}{\lam-x_i}\rp 
          \Bigg)\frac{\pho (\om)}{\pho (\lam)} \nonumber\\
&=& \frac{\partial}{\partial p}  \int\limits^{x_{2j}}_{x_{2j+1}}d\lam 
    \frac{1}{p-\lam}\frac{\pho (p)}{\pho (\lam)} 
 \ + \ \frac{1}{2}\sum_{i=1}^{2s}\dxi  M_i^{(1)} K_{i,j}  \ ,
\eeql
where now the contours have been deformed back to infinity. The first term
can be further evaluated, using partial integration,
\beql
\frac{\partial}{\partial p}  \int\limits^{x_{2j}}_{x_{2j+1}}d\lam 
    \frac{1}{p-\lam}\frac{\pho (p)}{\pho (\lam)} 
  &=& \frac{1}{2}\int\limits^{x_{2j}}_{x_{2j+1}}d\lam 
  \frac{1}{p-\lam}\sum_{i=1}^{2s}\lp\frac{1}{\lam-x_i}-\frac{1}{p-x_i}\rp 
   \frac{\pho (p)}{\pho (\lam)} \nonumber\\
  &=& \frac{1}{2}\sum_{i=1}^{2s}K_{i,j}\ \phi_i^{(1)}(p) \ ,
\eeql
which leads to the final form of eq. (\ref{dx}).

\sect{The planar two-loop correlator} \label{B}

\indent

The planar two-loop correlator $W_0(p,q)$ with $s$ cuts is derived by applying
$\dV\scr (p)$ from eq. (\ref{dVsum}) to the result for $W_0(p)$ eq. (\ref{W0}),
\beql
W_0(p,q) &=& \frac{1}{2}\frac{1}{(q-p)^2}\frac{\pho (q)}{\pho (p)}
             \ +\ \frac{1}{4}\frac{1}{(q-p)}\frac{1}{\pho (p)}
              \sum_{i=1}^{2s}\phi_i^{(1)}(q) \nonumber\\
           &&   \ - \frac{1}{2}\frac{1}{(q-p)^2}
              \ +\ \frac{1}{4}\frac{1}{\pho (p)}\sum_{i=1}^{2s}\frac{1}{p-x_i}
               M_i^{(1)} \frac{dx_i}{dV}(q) \nonumber\\
          &=& \frac{1}{4}\frac{\pho (q)}{\pho (p)} \lp
              \frac{2}{(q-p)^2}+\frac{1}{q-p}\sum_{i=1}^{2s}\frac{1}{p-x_i}
              +\sum_{i=1}^{2s}\sum_{l=0}^{s-2}\frac{\al_{i,l}\ q^l}{p-x_i} \rp 
               \nonumber\\
            &&- \frac{1}{2}\frac{1}{(q-p)^2} \ \ ,\label{W0pq}
\eeql
where the result for the $\frac{dx_i}{dV}\scr (q)$ eq. (\ref{dxsol}) 
has been used. Although
this is in principle the solution for arbitrary $s$, it would be nice to
have it in a form, where the analyticity properties, which follow from
its definition, are obviously fulfilled. Namely from eq. (\ref{Wdef})
\beq
W_0(p,q) \ =\ \dV (p)W_0(q) \ =\ \dV (p)\dV (q) F_0 \ ,
\eeq
it is clear, that it must satisfy
\begin{itemize}
\item symmetry $W_0(p,q)=W_0(q,p)$
\item analyticity like $W_0(p)\sim \pho (p)$, no higher poles
\item asymptotic $\lim_{p\to\infty}W_0(p,q)\sim {\cal O}(\frac{1}{p^2})$
\item regularity $\lim_{q\to p}W_0(p,q)=W_0(p,p)$
\item $-\ \frac{1}{(p-q)^2}=\lim_{\eps\to 0}\big(W_0(p+i\eps ,q))
 +W_0(p-i\eps ,q)\big)$
\end{itemize}
The last property can be derived from taking the derivative $\dV\scr (p)$
of eq. (\ref{ansatz}), which may be rephrased as 
\beq
\Vp (q) \ =\ 2\ \mbox{Re}\big( W_0(q)\big) \ \ , \ q\in \sigma \ \ .
\eeq
This last point is clearly satisfied by eq. (\ref{W0pq}), whereas the others 
are not obvious. In the case of the two-cut solutions however, where the 
$\al_{i,0}$ are explicitly known, 
$W_0(p,q)$ can be cast into a form, where all
points are seen to be true immediately. Inserting eq. (\ref{alf}) into eq.
(\ref{W0pq}) and using the identity
\beql
0&=& 2\prod_{i=1}^4(p-x_i)+(q-p)^2\Big( (p-x_2)(p-x_3)+(p-x_1)(p-x_4)\Big) 
                                                              \nonumber\\
&+&(q-p)\sum_{i=1}^4\prod_{j\neq i}(p-x_j)
                      \ -\ (p-x_1)(p-x_4)(q-x_2)(q-x_3)\nonumber\\
&-&(q-x_1)(q-x_4)(p-x_2)(p-x_3)
\eeql
leads to the planar two-loop correlator with two cuts,
\beql
W_0(p,q)&=&\frac{1}{4(p-q)^2}\Bigg(
  \sqrt{\frac{(p-x_1)(p-x_4)(q-x_2)(q-x_3)}{(p-x_2)(p-x_3)(q-x_1)(q-x_4)}}
\nonumber\\
&&\ \ \ \ \ \ \ \ \ \ \ \ 
   +\ \sqrt{\frac{(p-x_2)(p-x_3)(q-x_1)(q-x_4)}{(p-x_1)(p-x_4)(q-x_2)(q-x_3)}}
 \Bigg)\nonumber\\
      &+&\frac{1}{4}\ \frac{1}{\sqrt{\prod_{j=1}^4(p-x_j)(q-x_j)}}
         \frac{E(k)}{K(k)} f(\{x_i\})
     -\frac{1}{2(p-q)^2} \ ,
\label{W0pq2}
\eeql
with
\beql
f(\{x_i\})&=&       \frac{x_2x_3x_4(x_2-x_4)}{(x_1-x_4)(x_2-x_1)}
       + \frac{x_1x_3x_4(x_1-x_3)}{(x_1-x_2)(x_2-x_3)} \nonumber\\
       &+& \frac{x_1x_2x_4(x_4-x_2)}{(x_2-x_3)(x_3-x_4)}
       + \frac{x_1x_2x_3(x_1-x_3)}{(x_1-x_4)(x_3-x_4)}  \ . \label{W0pq2f}
\eeql
The same can be done for coinciding arguments, using eq. (\ref{dW0}), which 
reads
\beq
W_0(p,p)= \frac{1}{16}\sum_{i=1}^4\frac{1}{(p-x_i)^2}
        -\frac{1}{8}\sum^4_{i<j}\frac{1}{(p-x_i)(p-x_j)}
        +\frac{1}{4}\sum_{i=1}^{2s}\frac{\al_i}{p-x_i} \ .
\eeq
Taking the solution for the $\al_i$ eq. (\ref{alf}) it can be shown, that
\beql
\frac{1}{4}\sum_{i=1}^{2s}\frac{\al_i}{p-x_i}
&=&\frac{1}{4}\lp \frac{1}{(p-x_1)(p-x_4)}+\frac{1}{(p-x_2)(p-x_3)}\rp
  \nonumber\\
       &+&\frac{1}{4}\frac{1}{\prod_{j=1}^4(p-x_j)} \frac{E(k)}{K(k)} 
           f(\{x_i\}) \ ,
\label{W0pp2}
\eeql
which leads to the final form of eq. (\ref{Wpps}).
It can be checked now, that taking the limit $q\to p$ of eq. (\ref{W0pq2})
will give back the regular expression $W_0(p,p)$ as it should.

\sect{Elliptic integrals} \label{C}

\indent

The elliptic integrals defined in eq.(\ref{Ki}) can be expressed by
elementary functions and three fundamental integrals, namely the 
complete elliptic integrals of the first, second and third kind
\beql
K(k) &\equiv& \int\limits_0^1 dt \frac{1}{\sqrt{(1-t^2)(1-k^2t^2)}} \ ,\ \
E(k) \ \equiv\ \int\limits_0^1 dt \sqrt{\frac{1-k^2t^2}{1-t^2}}\ ,\nonumber\\
\Pi (\alpha^2,k) &\equiv& \int\limits_0^1 dt
                \frac{1}{(1-\alpha^2t^2)\sqrt{(1-t^2)(1-k^2t^2)}} \ .
       \label{ell}
\eeql
With the help of the tables in \cite {Byrd} the $K_i$, $i=1,\ldots,4$ , 
are then given by
\beql
K_1 &\equiv& -\bI \sqrt{\frac{(x_2-\lam)(\lam-x_3)(\lam-x_4)}{x_1-\lam}} 
\nonumber\\
    &=& X \Big[ (-\al^6 + 2\al^4 - 2\al^4k^2 + \al^2k^2)E(k)
               -(k^2-\al^2)(\al^4 - 2\al^2 + k^2)K(k)              \nonumber\\
     &&+ \ (\al^8 - 4\al^6k^2 + 6\al^4k^2 - 4\al^2k^2 + k^4)\Pi(\al^2,k) \Big]
                                                                \ , \nonumber\\
K_2 &\equiv& -\bI \sqrt{\frac{(x_1-\lam)(\lam-x_3)(\lam-x_4)}{x_2-\lam}} 
\nonumber\\
    &=& X \Big[ (-\al^6 - 2\al^4 + 2\al^4k^2 + \al^2k^2)E(k)
              +(-\al^4 - 2\al^2 + 4\al^2k^2 - k^2)\cdot  \nonumber\\
     && \ \ \cdot (k^2-\al^2)K(k)
         +(\al^8 - 4\al^6 + 6\al^4k^2 - 4\al^2k^4 + k^4)\Pi(\al^2,k)\Big] 
                                                                \ , \nonumber\\
K_3 &\equiv& \bI \sqrt{\frac{(x_1-\lam)(x_2-\lam)(\lam-x_4)}{\lam-x_3}} 
  \nonumber\\
    &=& X \Big[ (3\al^6 - 2\al^4 - 2\al^4k^2 + \al^2k^2)E(k)
               +(k^2-\al^2)(3\al^4 - 2\al^2 - k^2)K(k)             \nonumber\\
    &&+ \ (-3\al^8 + 4\al^6 + 4\al^6k^2 - 6\al^4k^2 + k^4)\Pi(\al^2,k) \Big]
                                                          \ , \nonumber\\
K_4 &\equiv& \bI \sqrt{\frac{(x_1-\lam)(x_2-\lam)(\lam-x_3)}{\lam-x_4}} 
 \nonumber\\
    &=& X \Big[ (-\al^6 + 2\al^4 + 2\al^4k^2 - 3\al^2k^2)E(k)
               +(-\al^4 + 2\al^2 - 4\al^2k^2 + 3k^2)\cdot \nonumber\\
     &&\cdot (k^2-\al^2)K(k)
        +(\al^8 - 6\al^4k^2 + 4\al^2k^2 + 4\al^2k^4 - 3k^4)\Pi(\al^2,k) 
                                                    \Big] \nonumber\\
&& \label{Klsg}
\eeql
where
\beql
X \ &\equiv& \ -\frac{1}{4}
  \frac{(x_1-x_3)^{\frac{3}{2}}(x_2-x_4)^{\frac{7}{2}}}{(x_2-x_3)^2(x_3-x_4)}
                                                                \ , \nonumber\\
k^2   \ &\equiv& \ \frac{(x_1-x_4)(x_2-x_3)}{(x_1-x_3)(x_2-x_4)}\ ,\ \ \ \ \ \ 
\al^2 \ \equiv \ \frac{x_2-x_3}{x_2-x_4} \ .  \label{kal}
\eeql
The composition into $K(k)$, $E(k)$ and $\Pi(\al^2,k)$ 
is not unique because of possible
transformations of the modulus $k$ and the parameter $\al$. Using the explicit 
form of eq. (\ref{Klsg}), the coefficients $\al_i$ in the solution for the
$\frac{dx_i}{dV}\scr (p)$ eq. (\ref{dxi2}) can be shown to have the simple
form, that is given in eq. (\ref{alf}).

\end{appendix}

\newpage


\newcommand{\NP}[3]{{\it Nucl. Phys. }{\bf B#1} (#2) #3}
\newcommand{\PL}[3]{{\it Phys. Lett. }{\bf B#1} (#2) #3}
\newcommand{\PR}[3]{{\it Phys. Rev. }{\bf #1} (#2) #3}
\newcommand{\IMP}[3]{{\it Int. J. Mod. Phys }{\bf #1} (#2) #3}
\newcommand{\MPL}[3]{{\it Mod. Phys. Lett. }{\bf #1} (#2) #3}

\end{document}